\documentclass[twocolumn,flushbottom,showpacs,preprintnumbers,superscriptaddress,amsmath,amssymb]{revtex4}
\usepackage{graphicx}
\usepackage{dcolumn}
\usepackage{bm}
\usepackage{amssymb}
\usepackage{amssymb}
\usepackage{color}
\usepackage{subfigure}
\def\>{\rangle}

\begin{document}

\title{Feedback-induced nonlinearity and superconducting on-chip quantum optics}

\author{Zhong-Peng Liu}
\affiliation{Department of Automation, Tsinghua University,
Beijing 100084, P. R. China} \affiliation{Center for Quantum
Information Science and Technology, TNList, Beijing 100084, P. R.
China}\affiliation{CEMS, RIKEN, Saitama, 351-0198, Japan}
\author{Hui Wang}
\affiliation{Institute of Microelectronics, Tsinghua University,
Beijing 100084, P. R. China} \affiliation{Center for Quantum
Information Science and Technology, TNList, Beijing 100084, P. R.
China} \affiliation{CEMS, RIKEN, Saitama, 351-0198, Japan}
\author{Jing Zhang}\email{jing-zhang@mail.tsinghua.edu.cn}
\affiliation{Department of Automation, Tsinghua University,
Beijing 100084, P. R. China} \affiliation{Center for Quantum
Information Science and Technology, TNList, Beijing 100084, P. R.
China} \affiliation{CEMS, RIKEN, Saitama, 351-0198, Japan}
\affiliation{State Key Laboratory of Robotics, Shenyang Institute
of Automation Chinese Academy of Sciences, Shenyang 110016, China}
\author{\\Yu-xi Liu}
\affiliation{Institute of Microelectronics, Tsinghua University,
Beijing 100084, P. R. China} \affiliation{Center for Quantum
Information Science and Technology, TNList, Beijing 100084, P. R.
China} \affiliation{CEMS, RIKEN, Saitama, 351-0198, Japan}
\author{Re-Bing Wu}
\affiliation{Department of Automation, Tsinghua University,
Beijing 100084, P. R. China} \affiliation{Center for Quantum
Information Science and Technology, TNList, Beijing 100084, P. R.
China}\affiliation{CEMS, RIKEN, Saitama, 351-0198, Japan}
\author{Chun-Wen Li}
\affiliation{Department of Automation, Tsinghua University,
Beijing 100084, P. R. China} \affiliation{Center for Quantum
Information Science and Technology, TNList, Beijing 100084, P. R.
China}\affiliation{CEMS, RIKEN, Saitama, 351-0198, Japan}
\author{Franco Nori}
\affiliation{CEMS, RIKEN, Saitama, 351-0198, Japan}
\affiliation{Physics Department, The University of Michigan, Ann
Arbor, Michigan 48109-1040, USA}

\date{\today}

\begin{abstract}
Quantum coherent feedback has been proven to be an efficient way
to tune the dynamics of quantum optical systems and, recently,
those of solid-state quantum circuits. Here, inspired by the recent
progress of quantum feedback experiments, especially those in
mesoscopic circuits, we prove that superconducting circuit QED systems, shunted with a
coherent feedback loop, can change the dynamics of a
superconducting transmission line resonator, i.e., a linear
quantum cavity, and lead to strong on-chip nonlinear optical
phenomena. We find that bistability can occur under the
semiclassical approximation, and photon anti-bunching can be shown
in the quantum regime. Our study presents new
perspectives for engineering nonlinear quantum dynamics on a chip.
\end{abstract}

\pacs{42.50.-p, 42.65.Pc, 02.30.Yy, 85.25.-j}

\maketitle

\section{Introduction}\label{s1}
Feedback, which is the core of modern control
theory~\cite{Astrom}, has been applied to the control of quantum
dynamical systems~\cite{Wiseman,Dong,Altafini,Kerckhoff0} for over twenty years, after
Belavkin's pioneering work on quantum filtering and
control~\cite{Belavkin}. Now, it has been extensively studied for
various problems in quantum system, such as optical
squeezing~\cite{Wiseman2,JGough1,Iida0}, spin
squeezing~\cite{LKThomsen}, quantum state
stabilization~\cite{JWang,Stockton,Handel,Ibarcq}, quantum error correction and noise
suppression~\cite{DVitali,Ganesan,Jzhang,Fzhang,Qi,Vijay,CAhn_and_ACDoherty,Xue}, entanglement
control~\cite{JWang_and_Wiseman,Yamamoto,zLiu},
cooling~\cite{AHopkins_and_KJacobs,jzhangl,Greenberg,Woolley}, and rapid quantum
measurement~\cite{JohnComes,jzhangl2}.

There are mainly two different quantum feedback control methods:
the measurement-based
feedback~\cite{Wisman1,Mabuchi,Doherty,Mancini} and coherent
feedback~\cite{Wisman4,Lloyd,Yanagisawa,James,Gough}. In a typical
measurement-based feedback-control optical system, a probe field
usually transmits through the quantum system to be
controlled and then the information is extracted.
Afterwards, the extracted information is fed into a classical
controller which generates the desired control signal. This control
signal is then fed back to tune the dynamics of the controlled
system. The simplest measurement-based feedback control protocol
is the so-called direct Markovian feedback~\cite{Wisman1}, in
which the measurement output is directly feedback to steer the
dynamics of the controlled system. Both the time-delay in the
feedback loop and the filtering effects induced by the integral
components are omitted, which leads to the so-called Markovian
approximation. However, due to the inevitable time delay and
filtering effects in the feedback loop, such a simplification may
not be valid in various cases. To solve this problem, another
measurement-based feedback protocol, called ``Bayesian
feedback", was proposed~\cite{Doherty}, in which the measurement
output signal is fed into a classical state estimator, e.g., a
series of integral components, to estimate the state of the
controlled quantum system and then fed into a classical controller
to obtain a state-based feedback control. It is worth noting that
both direct Markovian feedback and Bayes feedback have been
experimentally demonstrated in optical cavity
systems~\cite{Riste,Sciarrino,Brakhane,Yonezawa,Gieseler,Berry0,Kubanek,Sayrin,Zhou}
and solid-state circuits~\cite{Vijay}.

Although great progress has been achieved for measurement-based
feedback, there are still many problems left to be solved. The
main open problems of the measurement-based feedback approaches
include: (1) the time scale of the general quantum dynamics is too
fast to be manipulated in real time by currently-available
classical controllers; and (2), more essentially, the back-action
brought by the quantum measurement keeps dumping entropy into the
system before the feedback attempts to reduce it. One possible way
to solve this is to avoid the introduction of the
measurement step and use a fully quantum feedback loop to control
the quantum system, which leads to a new feedback mechanism called
coherent feedback. The simplest way to introduce coherent feedback
is to couple directly the controlled quantum system with the
quantum controller~\cite{Lloyd}, which is called ``direct coherent
feedback". An alternative approach is the field-mediated coherent
feedback~\cite{Yanagisawa,James,Gough}, in which the controlled
quantum system and the quantum controller are connected by an
intermediate quantum field. The direction of the information flow
in the feedback loop is naturally determined by the propagation
direction of the quantum field, and thus it is easier to be
realized in
experiments~\cite{Nelson,Mabuchi1,Zhou0,Kerckhoff,Iida}.

The existing studies about coherent feedback were mainly
focused on linear quantum systems. This is because previous
studies on coherent feedback are mainly focused on quantum
optical systems~\cite{Jaksch}, where the nonlinear effects are
too weak to be observed. However, recent progress shows that
nonlinear quantum optical phenomena induced by the strong
interaction between photons and solid-state components
can be observed~\cite{Birnbaum,Fink,Bishop} in solid-state
systems, such as quantum dots, superconducting circuits, and
silicon-based waveguides~\cite{Politi,Matthews,Berry}. In our
previous study~\cite{J.Zhang}, we found that, different from
measurement-based feedback, quantum coherent feedback can induce
and amplify the quantum nonlinear effects, and then modulate
the dynamics of the controlled system. We call this ``quantum feedback
nonlinearization". However, in this study, the quantum nonlinear
effects are induced by the nonlinear dissipative coupling between
the controlled quantum system and the intermediate quantum field.
The quantum feedback loop is just linear.

In this paper, we propose a different nonlinear coherent feedback
control strategy and apply it to superconducting
circuits. The main difference between this strategy
and our previous study in Ref.~\cite{J.Zhang} is that here
a nonlinear component, i.e., a nonlinear superconducting
device~\cite{You,Hoffman,Nation,Tornes,Buluta0,Georgescu,Buluta1,Nation1,Youf0,Youf1,Xiang0}, is embedded
in the feedback loop, and
the coupling between the controlled systems and the intermediate
quantum field is linear. Such a design is easier to be realized in
experiments~\cite{Kerckhoff}.

This paper is organized as follows.
In Sec.~\ref{s2}, we summarize results from the quantum
input-output theory and the theory of coherent feedback-control
networks, which will be used here afterwards.
In Sec.~\ref{s3}, we present our design of
nonlinear coherent feedback systems in supercoducting quantum circuits,
and then analyze the dynamics of the controlled systems in the
semiclassical regime (strong-driving regime) to show
bistability in Sec.~\ref{s4}, and the quantum regime
(weak-driving regime) to show quantum nonlinear optical phenomena,
such as photon antibunching effects in Sec.~\ref{s5}. Conclusions
and discussions are given in Sec.~\ref{s6}.

\section{PRELIMINARIES}\label{s2}
\begin{figure}[h]
\centerline{
\includegraphics[width=9cm]{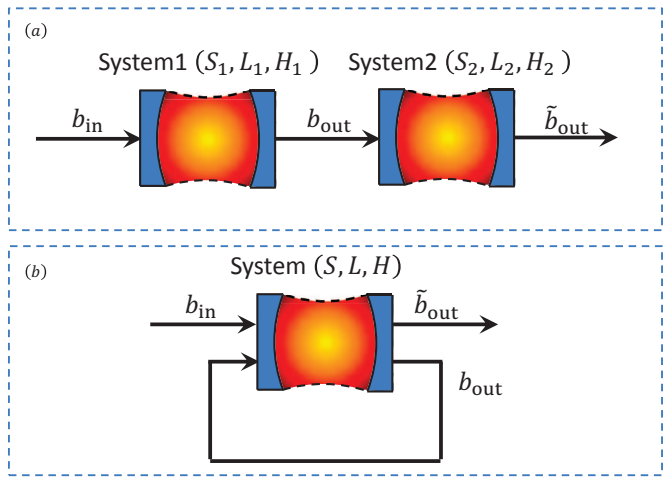}}
\caption{(Color online) Schematic diagram of the cascade system with
two cascaded-connected components in (a) and the feedback-control
systems in (b), which can be seen as the controlled system
cascaded-connected with itself.}\label{Fig of cascade}
\end{figure}

The basic model for a quantum input-output system can be presented
by a controlled system driven by an external bath, where the bath consists of
different modes which can be described by a continuum of harmonic
oscillators. We assume that $\hbar=1$ in the following
discussions. The Hamiltonian for such a system can be expressed as
\begin{equation}\label{Hamiltion SBI}
\begin{aligned}
H=&H_{\rm sys}+H_B+H_{\rm int}\\
H_B=&\int_{-\infty}^{+\infty}\omega b^{\dag}(\omega)b(\omega){\rm d}\omega\\
H_{\rm
int}=&i\int_{-\infty}^{+\infty}\left[\kappa(\omega)b^{\dag}(\omega)a-{\rm
h.c.}\right]{\rm d}\omega
\end{aligned}
\end{equation}
where $a$ is the annihilation operator of the system; and
$b^{\dag}(\omega)$, $b(\omega)$ are the creation and annihilation
operators of the bath mode with frequency $\omega$ satisfying
\begin{equation}
 \left[b(\omega),b^{\dag}(\tilde\omega)\right]=\delta\left(\omega-\tilde\omega\right).
\end{equation}
The commutator $\left[\cdot,\cdot\right]$ is defined as
$\left[A,B\right]=AB-BA$. $H_{\rm sys}$ is the free Hamiltonian of
the system, which interacts with the bath modes with coupling
operator $a$ and coupling strengths $\kappa(\omega)$. In the
interaction picture, Equation~(\ref{Hamiltion SBI}) can be
rewritten as
\begin{equation}
 H_{\rm eff}=H_{\rm sys}+i\int_{-\infty}^{+\infty}\left[\kappa(\omega)b^{\dag}(\omega)e^{i\omega t}a-{\rm h.c.}\right]{\rm
 d}\omega.
 \end{equation}
Under the Markov approximation, i.e., the coupling strength is
constant for all frequencies
$\kappa(\omega)=\sqrt{\gamma/2\pi}$, the Hamiltonian $H_{\rm eff}$
can be expressed as
\begin{equation}
H_{\rm eff}=H_{\rm sys}+i\left[b^\dag_{\rm in}L-L^\dag b_{\rm in}\right].
\end{equation}
$L=\sqrt{\gamma}a$ is the Lindblad operator induced by the
coupling between the system and the quantum field. Also,
\begin{equation}
b_{\rm in}(t)=\frac{1}{\sqrt{2\pi}}\int_{-\infty}^{+\infty}
b(\omega)e^{-i\omega t}{\rm d}\omega
\end{equation}
is defined as the input quantum field~\cite{Jacobs}.
\begin{figure}[h]
\centerline{
\includegraphics[bb=0 150 720 420,width=8cm,clip]{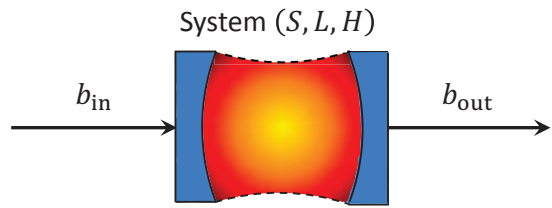}}
\caption{(Color online) Schematic diagram of the quantum input-output system
with the input field $b_{\rm in}$  and the output
field $b_{\rm out}$. Here, $S$, $L$, $H$ are, respectively, the scattering matrices, the Lindblad
operator, and the Hamiltonian of the input-output
system.}\label{Fig of input output}
\end{figure}
Let us consider a general input-output model as in Fig.~\ref{Fig
of input output}. The input field $b_{\rm
in}=[b_1(t),{\cdots},b_n(t)]^T$, where the superscript denotes transposition, satisfies the commutation
relations $[b_i(t),b^{\dag}_j(s)]={\delta}_{ij}(t-s)$, and transmits through a
beam-splitter described by an $n{\times}n$ unitary scattering
matrix $S$, such that $S^{\dag}S=SS^{\dag}=I$. The input field
interacts with the controlled system with the Hamiltonian $H$ which leads to the
dissipation channel represented by the Lindblad operator
$L=[L_1,{\cdots},L_n]^T$. The unitary evolution operator $U(t)$ of
the total system composed of the controlled system and the input field can be
described by the following quantum stochastic differential
equation~\cite{J.Zhang}:
 \begin{equation}\label{U}
 \begin{aligned}
 \frac{{\rm d}{U(t)}}{{\rm d}t}=&b^{\dag}_{\rm in}(t)(S-I)U(t)b_{\rm in}(t)+b^{\dag}_{\rm in}(t)LU(t)\\
 &-L^{\dag}SU(t)b_{\rm in}(t)-\left[\frac{1}{2}L^{\dag}L-iH\right]U(t),
 \end{aligned}
 \end{equation}
with initial condition $U(0)=I$, where $I$ is the identity
operator. The output field is defined by
\begin{equation}\label{input output relation}
b_{\rm out}(t)=U^{\dag}(t)SU(t)b_{\rm in}(t)+U^{\dag}(t)LU(t).
\end{equation}
It can be seen that the above quantum input-output system can be
fully determined by a set of operators $(S,L,H)$. For a quantum
Markovian cascaded system as shown in Fig.~\ref{Fig of cascade}(a),
in which the output field of the first component $(S_1,L_1,H_1)$
acts as the input field of the second system $(S_2,L_2,H_2)$, the
dynamics of the total system can be described by
\begin{equation}\label{SLH}
\left({S'},{L'},{H'}\right),
\end{equation}
with
\begin{equation*}
{S'}\equiv S_2S_1,\ {L'}\equiv L_2+S_2L_1,
\end{equation*}
\begin{equation*}
{H'}\equiv H_1+H_2+\frac{i}{2}\left(L_1^{\dag}S_2^{\dag}L_2-L_2^{\dag}S_2L_1\right).
\end{equation*}

In particular, we feed the output of the system $(S,L,H)$ back
and take it as the input of the same system to construct a direct
coherent-feedback network, as in Fig.~\ref{Fig of cascade}(b). From
Eq.~(\ref{SLH}), such a feedback network can be described
by~\cite{J.Zhang}:
\begin{equation}
\left(\tilde{S},\tilde{L},\tilde{H}\right),
\end{equation}
with
\begin{equation*}
\tilde{{S}}\equiv S^2,\ \tilde{{L}}\equiv L+SL,
\end{equation*}
\begin{equation*}
\tilde{{H}}\equiv H+\frac{i}{2}L^{\dag}(S^{\dag}-S)L.
\end{equation*}

\section{Nonlinear coherent feedback in superconducting circuits}\label{s3}

In Fig.~\ref{Fig of the schematic diagram of quantum
nonlinear feedback circuit} we present our proposed nonlinear
coherent feedback system using superconducting circuits. The controlled system
is a one-dimensional transmission line resonator (TLR) with
distributed inductor $L_s$ and capacitance $C_s$, which has two
input channels and two output channels. The TLR is driven by the
input field $b_{\rm in}$ through the dissipation channel
represented by the Lindblad operator $L=\sqrt{\gamma}a$. Also $a$ is
the annihilation operator of the quantum field in the TLR, and
$\gamma$ is the corresponding dissipation rate. The output field
of the TLR is then fed into a controller composed of another TLR
coupled to a dc-SQUID-based superconducting charge qubit,
 which is driven by either a current or voltage
source~\cite{G. Wendin,Beltran}. The quantum controller interacts
with the intermediate field via the dissipation channel with Lindblad
operator $L_c$. Afterwards, the output of the quantum controller
is fed back to act as the input field $\tilde{b}_{\rm in}$ of the
controlled system via the dissipation channel $L_f$ to close the coherent
feedback loop.

\begin{figure}[h]
\centerline{
\includegraphics[bb=90 0 600 540,width=8.5cm,clip]{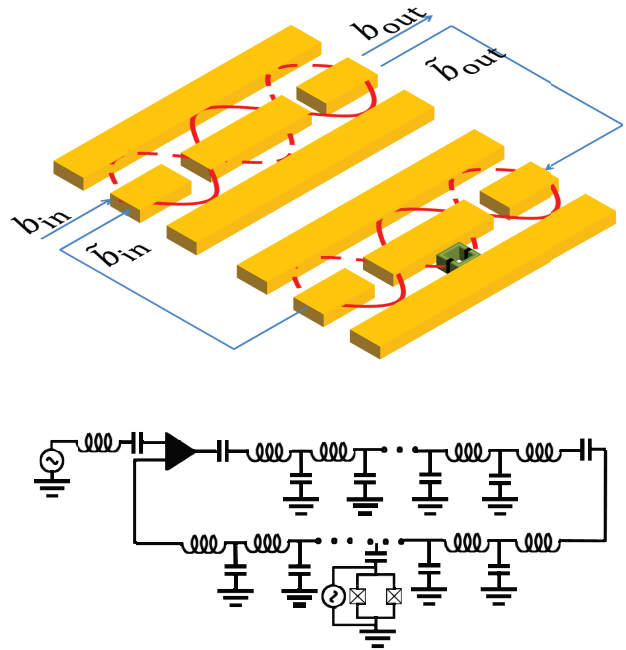}}
\caption{(Color online) Schematic diagram of the coherent
feedback network using superconducting quantum circuits. The controlled system is a TLR which can be considered
as a linear cavity. The controller in the feedback loop is another
TLR coupled with a dc-SQUID-based superconducting charge qubit which acts as a nonlinear
component.}\label{Fig of the schematic diagram of quantum
nonlinear feedback circuit}
\end{figure}
If we assume the inductance of the SQUID to be small, we can neglect
the magnetic energy of the circulating currents. Under the two-level
approximation, the Hamiltonian of the controller
can be expressed as~\cite{Liu1}
\begin{equation}
\begin{aligned}
 H_{qT}&=\frac{1}{2}\omega_0\sigma_z+\omega_c c^{\dag}c+ g\left(c\sigma_++c^{\dag}\sigma_-\right)\\
 &+\Omega\left(\sigma_+e^{-i\omega_1 t}+\sigma_-e^{i\omega_1 t}\right)+\varepsilon\left(c^{\dag}e^{-i\omega_2 t}+ce^{i\omega_2 t}\right),\label{Hamiltonian of dc SQUID}
 \end{aligned}
\end{equation}
where $c$ is the annihilation operator of the quantized electromagnetic
field in the TLR coupled to the dc-SQUID based superconducting charge qubit; $\Omega$ is the Rabi
frequency describing the interaction between the qubit and the classical field; $g$ is the coupling strength between
the qubit and the quantized electromagnetic
field; and $\varepsilon$ is the strength of the
driving field applied to the cavity mode.
If the Rabi frequency $\Omega$ is large enough, such that
$\Omega\gg g^2/\Delta_{qT}$, and in the large-detuning regime,
\begin{equation*}
\Delta_{qT}=\omega_c-\omega_0\gg g,
\end{equation*}
 we can write an
effective Hamiltonian which can be
reexpressed as~\cite{Liu1} (see derivations in Appendix A)
 \begin{equation}\label{effect H}
H_{\rm eff}=\omega_c
c^{\dag}c+\Omega\sigma_z+\frac{g^4}{2\Omega\Delta_{qT}^2}\left(c^{\dag}c\right)^2\sigma_z
\end{equation}
If the qubit is adiabatically placed in its ground state $|-\rangle$, the
effective Hamiltonian in Eq.~(\ref{effect H}) can be expressed as
\begin{equation}\label{nonlinear Hamiltonian}
 \begin{aligned}
H_{qT}^{\prime}&=\omega_c c^{\dag}c-\chi\left(c^{\dag}c\right)^2\\
&+\varepsilon\left(c^{\dag}e^{-i\omega_2 t}+ce^{i\omega_2
t}\right),
 \end{aligned}
\end{equation}
where $\chi={g^4}/\left({2\Omega\Delta_{qT}^2}\right)$. In the
$(S,L,H)$ notation, the quantized electromagnetic
field can be represented by:
\begin{equation}
(1, \sqrt{\kappa}{c},H_{qT}^{\prime}),
\end{equation}
where $\kappa$ is the decay rate of the quantized electromagnetic
field. The
controlled linear cavity (TLR) can be described as
$(1,\sqrt{\gamma}{a},\omega_s{a^{\dag}a})$, where $\gamma$ and $a$
are the decay rate and the annihilation operator of the cavity,
respectively.

The total coherent feedback system can be considered as a
cascade system of the subsystem
$(1,\sqrt{\gamma}{a},\omega_s{a^{\dag}a})$, the quantum controller
$(1, \sqrt{\kappa}{c},H_{qT}^{\prime})$, and the subsystem
$(1,\sqrt{\gamma_f}{a},\omega_s{a^{\dag}a})$, which can be written
as
\begin{equation}
 \begin{aligned}
 (S'',L'',H''),
 \end{aligned}
 \end{equation}
 with
 \begin{equation*}
S''=1,\ L''=\sqrt{\kappa}{c}+(\sqrt{\gamma}+\sqrt{\gamma_f}){a},
\end{equation*}
 \begin{equation*}
H''=\omega_s{a^{\dag}a}+H_{qT}^{\prime}+\frac{i}{2}(\sqrt{\kappa\gamma}-\sqrt{\kappa\gamma_f})(a^{\dag}c-c^{\dag}a).
\end{equation*}
In the rotating reference frame with unitary transformation
$V(t)=\exp\left[i\left(\omega_2{c^{\dag}}c+\omega_2{a^{\dag}}a\right)t\right]$,
the total Hamiltonian can be represented as
 \begin{equation}\label{total Hamiltonian}
 \begin{aligned}
 {H}_{\rm tot}=&\Delta_s{a^{\dag}}a+{\Delta}c^{\dag}c-\chi\left(c^{\dag}\right)^2c^2-\varepsilon(c^{\dag}+c)\\
 &+\frac{i}{2}(\sqrt{\kappa\gamma}-\sqrt{\kappa\gamma_f})(a^{\dag}c-c^{\dag}a),
\end{aligned}
 \end{equation}
where
 \begin{equation*}
\Delta_s=\omega_{s}-\omega_2,\ \Delta=\omega_c-\omega_2,
\end{equation*}
are the detuning frequencies.

With the decrease of the strength of the external driving field,
different optical phenomena can be observed. In the strong-driving
semi-classical regime, semiclassical nonlinear
bistability effects can be found, while in the weak-driving quantum regime,
quantum nonlinear phenomena such as anti-bunching can be observed.

\section{Nonlinear on-chip optics by coherent feedback: semi-classical regime}\label{s4}

We first consider the case when the external field imposed on the
controller is strong, where we can observe semiclassical
nonlinear phenomena such as optical bistability. Optical
bistability is a typical nonlinear phenomenon which has
been observed in various
systems~\cite{Gibbs,Rempe,Sauer,Gupta,Brennecke,Xiao}. It has also
been demonstrated that quantum feedback can modulate such kinds of
nonlinear effects. For example, the recent
experiment~\cite{Kerckhoff} showed that two coupled
superconducting tunable Kerr cavities (TKCs) connected in a
feedback configuration can demonstrate semiclassical on-chip
nonlinear optical phenomena. The reflected phase from the TKC is a
nonlinear function of the driving amplitude and thus the TKC acts as a typical
nonlinear component, which means that both the controlled system
and the controller are nonlinear systems. In contrast to this design,
in our system, as shown in Fig.~\ref{Fig of the schematic diagram of
quantum nonlinear feedback circuit}, the controlled system is a
linear system and the controller is nonlinear. We now want to
show how the nonlinear controller modulates the controlled linear
dynamics, making the controlled system nonlinear.

\begin{figure}[ht]
\centerline{
\includegraphics[bb=43 245 515 580, width=8.5cm,clip]{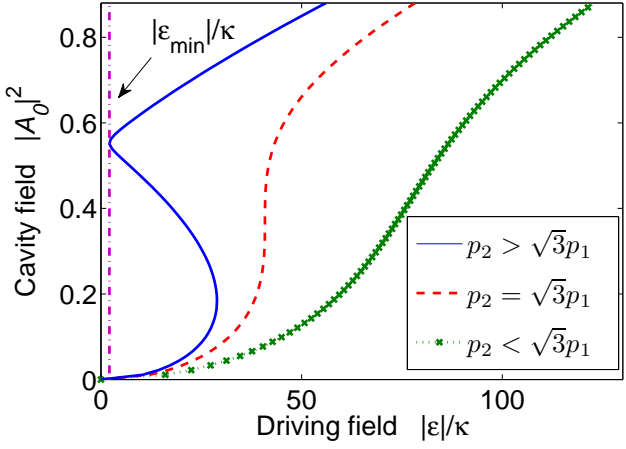}}
\caption{(Color online) Magnitude of the steady field $|A_0|^2$ in the
controlled cavity versus the intensity of the driving field
$|\varepsilon|$, with system parameters: $\Delta_s/{2\pi}=100$
MHz, $\Delta/{2\pi}=4.9$ MHz, $\chi/{2\pi}=10$ MHz,
$\gamma/{2\pi}=6$ MHz, $\gamma_f/{2\pi}=8$ MHz and $\kappa/2\pi=3$
MHz. In the parameter regime $p_2>\sqrt{3} p_1$, we observe
a hysteresis curve, indicating bistability. In the parameter regime $p_2\leq\sqrt{3} p_1$, bistability disappears.}\label{Fig of bistability}
\end{figure}
Based on the Hamiltonian in Eq.~(\ref{total Hamiltonian}),the
Heisenberg-Langevin equations of the total system can be described
by:
\begin{eqnarray}
 \dot{a}&=&-i\Delta_s{a}-\frac{1}{2}(\sqrt{\gamma}+\sqrt{\gamma_f})^2a-\sqrt{\kappa\gamma_f}c\nonumber\\
 &&-(\sqrt{\gamma}+\sqrt{\gamma_f})b_{\rm in},\label{lang of a} \\
 \dot{c}&=&-i\Delta{c}+i\chi(2c^{\dag}cc)-\frac{\kappa}{2}c-\sqrt{\kappa\gamma}a\nonumber\\
 &&+i\varepsilon
    -\sqrt{\kappa}b_{\rm in},\label{lang of c}
\end{eqnarray}
where $b_{\rm in}$ is the vacuum field with zero mean value
$\langle{b_{\rm in}(t)}\rangle=0$ and delta correlation
$\langle{b_{\rm
in}(t)b^{\dag}_{\rm in}(t^{\prime})}\rangle=\delta(t-t^{\prime})$,
where $\left\langle\cdot\right\rangle$ represents the average over
the equilibrium state of the environment. The input-output
relation of the total system can be written as
\begin{equation}
b_{\rm out}=\left(\sqrt{\gamma_f}+\sqrt{\gamma}\right)a+\sqrt{\kappa}c+b_{\rm
in}.
\end{equation}
Using the mean field approximation, the time evolutions of the
mean values of the operators $a$ and $c$ can be given
by~\cite{Hui Wang}:
\begin{eqnarray}\label{syeady of a and c}
\frac{{\rm d}\langle{a}\rangle}{{\rm
d}t}&=&-i\Delta_s\langle{a}\rangle-\frac{1}{2}(\sqrt{\gamma}+\sqrt{\gamma_f})^2\langle{a}\rangle-\sqrt{\kappa\gamma_f}\langle{c}\rangle,\nonumber\\
\frac{{\rm d}\langle{c}\rangle}{{\rm
d}t}&=&-i\Delta\langle{c}\rangle+i\chi(2{\langle{c^{\dag}}\rangle}{\langle{c}\rangle}^2)-\frac{\kappa}{2}\langle{c}\rangle\nonumber\\
&&-\sqrt{\kappa\gamma}\langle{a}\rangle+i\varepsilon.
\end{eqnarray}

By assuming that the
steady values of $\langle{a}\rangle$ and $\langle{c}\rangle$ are
$A_0$ and $C_0$, we can obtain $A_0$ and $C_0$ by the following
equations:
\begin{eqnarray}
&4\chi^2{\left|C_0\right|}^6-4p_2\chi{\left|C_0\right|^4}+(p_1^2+p_2^2){\left|C_0\right|^2}=\left|\varepsilon\right|^2,&\label{kerr}\nonumber\\
&\left|A_0\right|^2=\frac{4{\kappa{\gamma_f}}}{4\Delta_s^2+(\sqrt{\gamma}+\sqrt{\gamma_f})^4}
\left|C_0\right|^2,&
\end{eqnarray}
with
\begin{eqnarray*}
p_1&=&\frac{\kappa}{2}-\frac{2\kappa\sqrt{\gamma\gamma_f}(4\sqrt{\gamma}+\sqrt{\gamma_f})^2}{4\Delta_s^2+(\sqrt{\gamma}+\sqrt{\gamma_f})^4},\\
p_2&=&\Delta+\frac{4\kappa\sqrt{\gamma\gamma_f}\Delta_s}{4\Delta_s^2+(\sqrt{\gamma}+\sqrt{\gamma_f})^4}.
\end{eqnarray*}
When $p_2>\sqrt{3}p_1$, Eq.~(\ref{kerr}) has two stable
solutions corresponding to a local maximum and a local
minimum, respectively, which means that the system is in a
bistable regime. Indeed, the maximum of $\left|C_0\right|^2$ can
be found by $\partial{\left|C_0\right|^2}/\partial{\Delta}=0$,
which leads to
\begin{equation}\label{N/D}
-2\chi \left|C_0\right|^2 + p_2 = 0.
\end{equation}
Substituting Eq.~(\ref{N/D}) into Eq.~(\ref{kerr}), we can obtain
\begin{equation}\label{ep}
\left|\varepsilon\right|^2=p_1^2 \left|C_0\right|^2,
\end{equation}
by which we can find the strength of the driving field that
maximizes the intracavity intensity for fixed $\Delta$. Additionally,
we also know that the bifurcation line with the current drive and
detuning parameter which provides the boundary between the single-solution
and bistable-solution regions is located where the susceptibility
$\partial{C}/\partial{\Delta}$ diverges, where $C=|C_0|^2$. The bistability occurs
when
$\partial{\Delta}/\partial{C}=\partial^2{\Delta}/\partial{C^2}=0$
holds. This condition leads to $12\chi^2C^2-8p_2\chi
C+p_1^2+p_2^2=0$ and $3\chi^2 C-p_2\chi=0$. Thus, it is easy to
find that the critical point of bistability $p_2^2=3p_1^2$ (from
which we find that $\left|p_2\right|>\sqrt{3}\left|p_1\right|$
and $C>\sqrt{3}\left|p_1\right|/{\chi}$) should be satisfied to
remain in the stable region. Substituting these into Eq.~(\ref{ep}), we relate
$p_1$ and the driving field intensity $\varepsilon$ in
the stable region with a maximum intracavity intensity as
\begin{equation}
 \left|\varepsilon\right|^2>\frac{\sqrt{3}p_1^3}{\chi}.
 \end{equation}
From Fig.~\ref{Fig of bistability}, we find there is a two steady-state
relation between the steady-state solution for the controlled system and
driving intensity, where
$\left|\varepsilon\right|^2>\sqrt{3}p_1^3/\chi$. And the two steady-state
solutions can be lost when
$\left|\varepsilon\right|^2\leq\sqrt{3}p_1^3/\chi$.

\section{Nonlinear on-chip optics by coherent feedback: quantum regime}\label{s5}


We now consider the case when the external field imposed on the
controller is weak. In this case, the controlled system is in the
full quantum regime.~Under the Markovian approximation, the evolution
of the total system can be described by the following master
equation~\cite{Puri}:
\begin{equation}\label{rho}
\frac{{\rm d}\rho}{{\rm d}t}=-{i}[H_{\rm
tot},\rho]+\frac{1}{2}(2L_{\rm tot}\rho{L^{\dag}_{\rm
tot}}-{L^{\dag}_{\rm tot}}L_{\rm tot}\rho-\rho{L^{\dag}_{\rm
tot}}L_{\rm tot}),
\end{equation}
where the Lindblad operator $L_{\rm tot}$ can be represented as
\begin{equation*}
L_{\rm tot}=(\sqrt{\gamma}+\sqrt{\gamma_f})a+\sqrt{\kappa}c.
\end{equation*}
From Eq.~(\ref{rho}), we obtain the steady state $\rho_{\rm ss}$
by letting $d\rho/dt=0$, and thus we can calculate the following
normalized second-order correlation function of the controlled
system:
\begin{equation}\label{g2}
 {g}^{(2)}(0)=\frac{\langle(a^{\dag})^2a^2\rangle}{\langle{a}^{\dag}a\rangle^2}\equiv\frac{{\rm Tr}(\rho_{\rm ss}(a^{\dag})^2a^2)}{\left[{\rm
 Tr}(\rho_{\rm ss}a^{\dag}a)\right]^2},
\end{equation}
which provides the information of the photon number statistics of
the single-mode cavity field in the TLR. Let us define a new parameter ${K}=\Delta/\chi+1$, which can
be tuned by adjusting the detuning $\Delta$, where ${K}=1$ and ${K}=2$
mean that the system is in the single-photon and two-photon
blockade regimes~\cite{Adam}.
\begin{figure}[h]
\centerline{
\includegraphics[bb=43 251 515 580,width=8.5cm,clip]{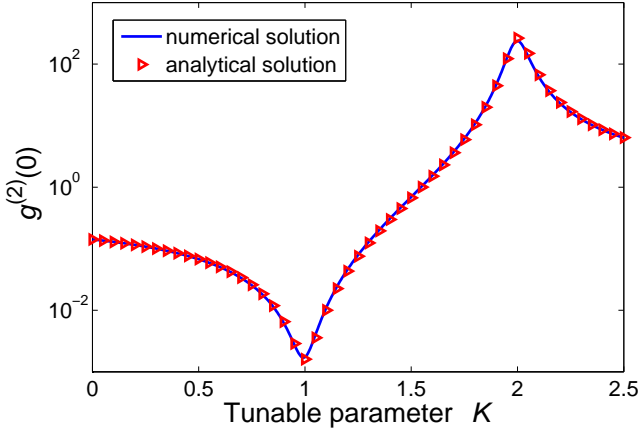}}
\caption{(Color online). The numerical and analytical solutions
for the second-order correlation function $g^{(2)}(0)$ in Eq.~(\ref{g2}) versus the
tunable parameter ${K}$. The system parameters for this simulation
are: $\chi=10$ MHz, $\Delta_s=50$ MHz, $\gamma/2\pi=2$ MHz,
$\gamma_f/2\pi=2.5$ MHz, $\kappa=1$ MHz, and $\varepsilon=0.1\kappa$.}\label{Fig of
numerical solution and analytical solution}
\end{figure}
We compare the numerical and analytical solutions of ${g}^{(2)}(0)$
in Fig.~\ref{Fig of numerical solution and analytical solution}. The analytical
solution of $g^{\left(2\right)}\left(0\right)$ is obtained in the
following way. From Eq.~(\ref{g2}), we have~\cite{Xu}
\begin{equation}\label{g2p}
 g^{(2)}(0)=\frac{\sum\limits_n n(n-1)P_n}{\left(\sum\limits_n
 nP_n\right)^2},
\end{equation}
where $P_n$ represents the probability with $n$ photons. In the
weak-driving limit, i.e., $\varepsilon\to0$, it can be shown that
$P_n\gg P_{n+1}$ for $n \geq 2$. Thus, we can omit the probability
for three or more photons. In this case, the steady state of the
system can be expressed as:
\begin{equation}
 \left|\psi\right\rangle=\sum\limits_{n_p=0}^2\sum\limits_{n_c=0}^2
 C_{n_p,n_c}\left|n_p,n_c\right\rangle,
\end{equation}
where $\left|n_p\right\rangle$ and $\left|n_c\right\rangle$ are
the states of the controlled system and the controller
respectively. In order to find the coefficients $C_{n_p,n_c}$ for
the steady state, we introduce the complex Hamiltonian by letting
\begin{equation*}
\Delta_s\to\Delta_s-i(\sqrt\gamma+\sqrt\gamma_f)^2/2,
\end{equation*}
\begin{equation*}
\Delta\to\Delta-i\kappa/2
\end{equation*}
to represent the dissipation effects.
Then, the steady state can be found via ${\rm
d}\left|\psi(t)\right\rangle/{\rm d}t=0$. The probability of the
$n_p$-th occupation number can be expressed as
\begin{equation*}
P_{n_p}=\sum\limits_{n_c}\left|C_{n_p,n_c}\right|^2.
\end{equation*}
Thus, in
the weak-driving limit, $g^{(2)}(0)$ can be described
by (see derivations in Appendix B)
\begin{equation}\label{analytical}
 g^{(2)}(0)=\frac{\left|\Delta_s+(K-2)\chi-\frac{i}{2}\gamma_a\right|^2\left|(K-1)\chi-\frac{i}{2}\kappa
\right|^2}{\left|\left((K-2)\chi-\frac{i}{2}\kappa\right)\left(\Delta_s+(K-1)\chi-\frac{i}{2}\gamma_a\right)\right|^2},
\end{equation}
with
\begin{equation*}
\gamma_a=(\sqrt\gamma+\sqrt\gamma_f)^2+\kappa.
 \end{equation*}
In Fig.~\ref{Fig of numerical solution and analytical solution},
the analytical result for $g^{(2)}(0)$ fits well with the
numerical solution obtained by the few photon truncation in the
weak-driving regime. It can be shown that
 \begin{equation}
 g^{(2)}(0)=\frac{4\kappa^2\left(\Delta_s-\chi\right)^2+\kappa^2\gamma_a^2}{4\kappa^2\Delta_s^2+16\chi^2\Delta_s^2+\left(4\chi^2+\kappa^2\right)\gamma_a^2},
 \end{equation}
when $K=1$. When the dissipation coefficients $\kappa$, $\gamma$,
and $\gamma_f$ are far less than the Kerr nonlinear coefficient
$\chi$ and the detunning frequency $\Delta_s$, we have
$g^{(2)}(0)<1$ which leads to the single-photon blockade.
 \begin{figure}[h]
\centerline{
\includegraphics[bb=43 185 520 655,width=8.8cm,clip]{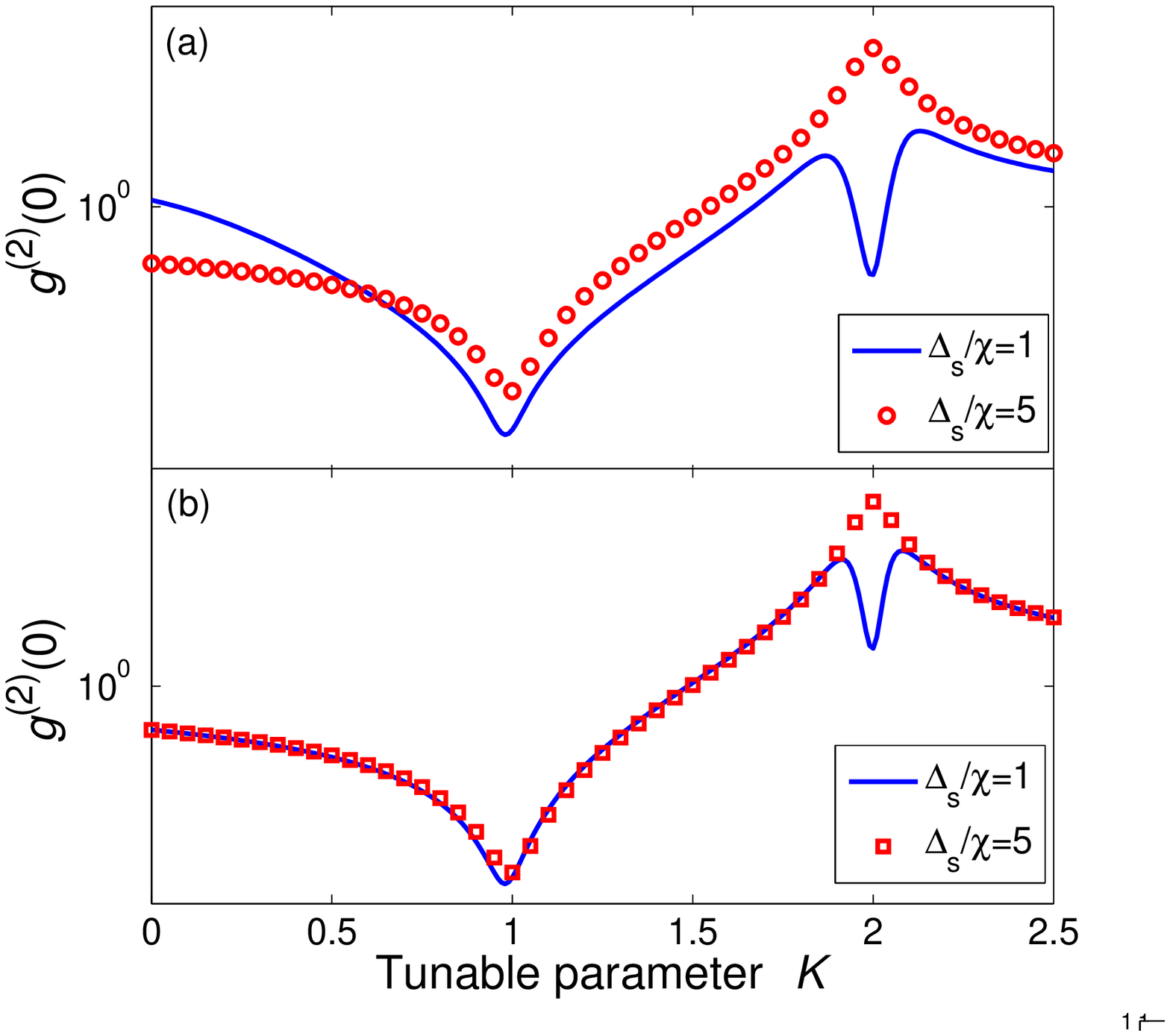}}
\caption{(a) $g^{(2)}(0)$ for the controlled TLR and (b)
$g^{(2)}(0)$ for the controller versus the parameter
$K=\Delta/\chi+1$. The red dotted curve and the blue curve
correspond to different detuning frequencies $\Delta_s=50$ MHz and
$\Delta_s=10$ MHz. The other system parameters are: $\chi=10$ MHz
$\gamma/2\pi=2$ MHz, $\gamma_f/2\pi=2.5$ MHz, $\kappa=1$
MHz, and $\varepsilon=0.1\kappa$.}\label{Fig of second order function}
\end{figure}
We plot the curve of $g^{(2)}(0)$ for the controlled single-mode cavity in the TLR in
Fig.~\ref{Fig of second order function}(a). We find that there
is a minimum point of $g^{(2)}(0)$ at $K=1$ for each curve, which
means that photon antibunching occurs. This typical nonlinear
quantum phenomenon observed in the linear-controlled single-mode cavity in the TLR is
induced by the Kerr nonlinearity in the controller by coherent
feedback.

However, the curves for different detuning frequencies $\Delta_s$
are different at $K=2$. For $\Delta_s/\chi=1$, we can observe that
$g^{\left(2\right)}\left(0\right)<1$ which means that photon
blockade occurs. In this case, it can be shown from $K=2$
and $\Delta_s/\chi=1$ that $\omega_s=\omega_c$, which means that
the controller and the controlled TLR resonate with each other.
When the resonance occurs, if one photon enters the controlled
TLR, this photon will be transmitted to the controller from the
feedback loop and then transmitted back. It will block the next
photon to enter the controlled TLR. In Fig.~\ref{Fig of second
order function}(b), we can find for the curve with
$\Delta_s/\chi=5$ that $g^{\left(2\right)}\left(0\right)>1$ at
$K=2$, which means two photons resonating. This corresponds to a
transparency effect.
\begin{figure}[h]
\centerline{
\includegraphics[bb=43 240 515 580,width=8.5cm,clip]{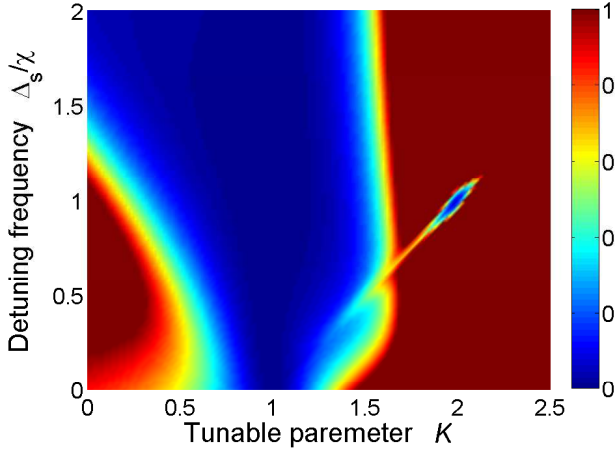}}
\caption{(Color online). Second-order correlation function versus
$K$ and $\Delta_s$, with $\chi=10$ MHz,
$\gamma/2\pi=2$ MHz, $\gamma_f/2\pi=2.5$ MHz, $\kappa=1$
MHz and $\varepsilon=0.1\kappa$. Note that $g^{(2)}(0)<1$ occurs when $\Delta_s-(K-1)=0$, namely, $\omega=\omega_s$
and $K=1$ with any $\Delta_s$. }\label{Fig of Second-order
correlation function depend on n and delta}
\end{figure}

In Fig.~\ref{Fig of Second-order correlation function depend on n
and delta}, we show how $g^{(2)}(0)$ depends on the parameters $K$
and $\Delta_s$. The photon antibunching occurs at $K=1$ with any
$\Delta_s$. However, at $K=2$ it can only be observed when
$\Delta_s/\chi=1$, which means that the controlled TLR and the
controller are resonant with each other.

\begin{figure}[h]
\centerline{
\includegraphics[bb=43 185 510 645,width=8.8cm,clip]{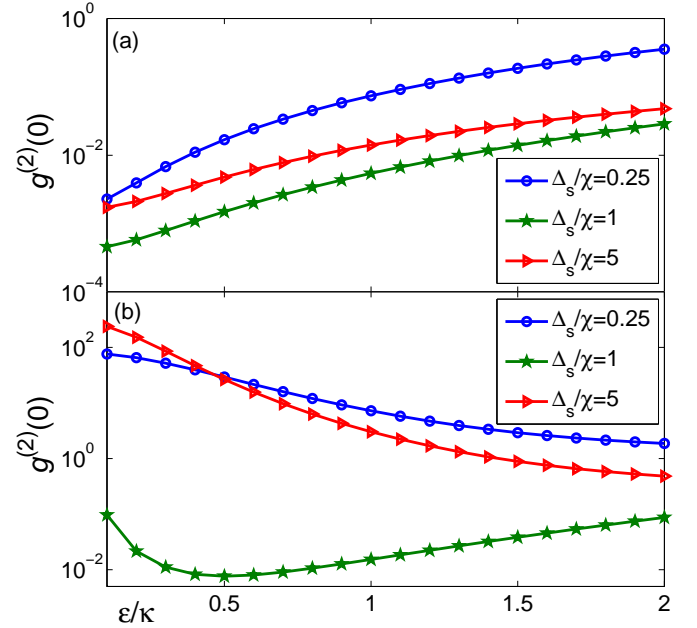}}
\caption{(Color online). Second-order correlation function versus the normalized driving
strength $\varepsilon/\kappa$, for different values of the detuning frequency $\Delta_s$. Here the
system parameters are: $\chi=10$ MHz, $\gamma/2\pi=2$ MHz,
$\gamma_f/2\pi=2.5$ MHz, and $\kappa=1$ MHz. (a) for $K=1$ and (b) for $K=2$.}\label{Fig of
Second-order correlation function depend on ep}
\end{figure}

In Fig.~\ref{Fig of Second-order correlation function depend on
ep}, we also study how $g^{(2)}(0)$ depends on the strength of the
driving field. In Fig.~\ref{Fig of Second-order correlation
function depend on ep}(a), we find that $g^{(2)}(0)\to 1$ when increasing
the strength of the driving field. We can also
observe that $g^{(2)}(0)$ is minimized when the resonance occurs,
i.e., $\Delta_s/\chi=1$. Similar to
Fig.~\ref{Fig of Second-order correlation function depend on
ep}(a), in Fig.~\ref{Fig of Second-order correlation function depend on
ep}(b) we find that all the curves converge to $1$ when increasing
the strength of the driving field. However, we can
obtain a concave curve when the resonance occurs, i.e.,
$\Delta_s/\chi=1$.

\begin{figure}[h]
\centerline{
\includegraphics[bb=43 245 550 597,width=8.8cm,clip]{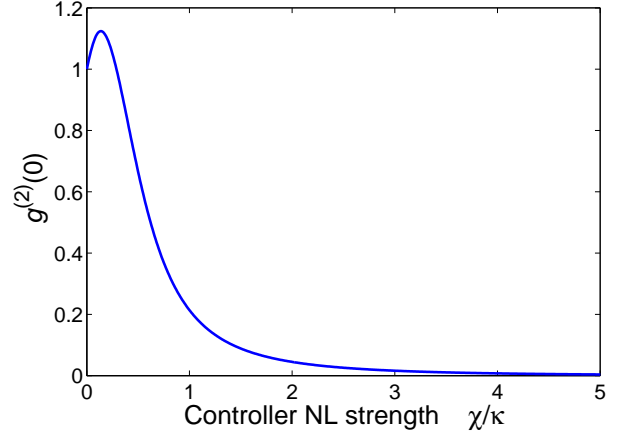}}
\caption{(Color online). Second-order correlation function versus the nonlinear strength $\chi$ of the controller,  where the parameters of the system are: $K=1$, $\gamma/2\pi=2\ \rm MHz$, $\gamma_f/2\pi=2.5$ MHz, $\kappa=1$ MHz and $\varepsilon=0.1\kappa$.}\label{Fig of Second-order correlation function depend on x}
\end{figure}
The generated nonlinear effects in the controlled single-mode field in the TLR can be
enhanced by increasing the nonlinear strength of the controller,
and the nonlinear strength $\chi$ of the controller can be tuned
by adjusting the detuning frequency $\Delta_{qT}$. Thus, we can
enhance the generated nonlinear effects in the controlled TLR by
tuning $\Delta_{qT}$. Indeed, from Fig.~\ref{Fig of Second-order
correlation function depend on x}, we can find that the antibuching
effects in the controlled TLR are enhanced by decreasing the
detuning $\Delta_{qT}$.

\section{Conclusions}\label{s6}
In summary, we presented a nonlinear coherent feedback-control
system in superconducting circuits in which a controlled linear
transmission line resonator is modulated by another transmission
line resonator coupled to a dc-SQUID-based superconducting charge qubit
 in the feedback loop. Such
a design changes the linear dynamics of the controlled resonator
and makes it nonlinear. This nonlinear coherent feedback-control
system is used to produce strong nonlinear on-chip optical
phenomena. In the semiclassical regime, we observe in simulations
bistable optical-type phenomena observed in previous optical
experiments, and give the condition to observe this bistability.

In
the quantum regime, we predict a photon antibunching
induced by nonlinear coherent feedback, which is believed to be a
typical quantum optics phenomenon that
violates the Cauchy-Schwartz inequality for classical light.
Our study shows that dynamics of linear system can be switched to
nonlinear one by using nonlinear coherent feedback in a controllable way.
We hope that our prediction of nonlinear coherent feedback in
the quantum regime can be verified experimentally in the near future.

\begin{center}
\textbf{ACKNOWLEDGMENTS}
\end{center}

ZPL would like to thank Dr.~X.W. Xu for helpful discussions. JZ
and RBW are supported by the NSFC under Grant Nos.61174084,
61134008, 60904034, and project supported by State Key Laboratory
of Robotics, Shenyang Institute of Automation Chinese Academy of
Sciences, China. YXL is supported by the NSFC under Grant
Nos.~61025022, 60836001. FN is partially supported by
the ARO, RIKEN iTHES Project, MURI Center for Dynamic
Magneto-Optics,  JSPS-RFBR contract
No.12-02-92100, Grant-in-Aid for Scientific Research (S), MEXT
Kakenhi on Quantum Cybernetics, and the JSPS via its FIRST
program.

\appendix
\section{qubit-induced nonlinearity}
If the detuning frequency $\Delta_{qT}=\omega_c-\omega_0$ between the qubit and the TLR is much lager than the intensity $g$ between the qubit and the TLR, by making the unitary transformation
 \begin{equation}
U_0=\exp\left[\frac{g}{\Delta_{qT}}\left(c\sigma_+-c^{\dag}\sigma_-\right)\right],
\end{equation}
the effective Hamiltonian can be described as
 \begin{equation}
 \begin{aligned}
H_{\rm eff}&=U_0HU_0^{\dag}\\
&\approx\left(\omega_c+\frac{g^2}{\Delta_{qT}}\sigma_z\right)c^{\dag}c+\frac{1}{2}\left(\omega_0+\frac{g^2}{\Delta_{qT}}\right)\sigma_z\\
&+\Omega\left(\sigma_+e^{-i\omega_1 t}+\sigma_-e^{i\omega_1 t}\right)
\end{aligned}
\end{equation}
In the rotating reference frame with frequency $\omega_1$ of the driving field, we let
\begin{equation}
U_1=\exp\left[-i\omega_1\frac{\sigma_z}{2}\right]t
\end{equation}
where $\omega_1=\omega_0+\frac{g^2}{\Delta_{qT}}$, the Hamiltonian can then be described by
\begin{equation}
H_{\rm eff}'=\omega_c c^{\dag}c+\Omega\sigma_x+\frac{g^2}{\Delta_{qT}}c^{\dag}c\sigma_z
\end{equation}
We set $\sigma_z\rightarrow\sigma_x$, and then
\begin{equation}
H_{\rm eff}''=\omega_c c^{\dag}c+\Omega\sigma_z+\frac{g^2}{\Delta_{qT}}c^{\dag}c\left(\sigma_+ +\sigma_-\right).
\end{equation}
If the Rabi frequency $\Omega$ satisfies the condition $\Omega\gg(g^2/\Delta_{qT})$, through the unitary transformation
 \begin{equation}
U_2=\exp\left[\frac{g^2}{2\Omega\Delta_{qT}}c^{\dag}c\left(\sigma_+-\sigma_-\right)\right],
\end{equation}
the Hamiltonian can be re-expressed by
 \begin{equation}
H_{\rm eff}'''=\omega_c c^{\dag}c+\Omega\sigma_z+\frac{g^4}{2\Omega\Delta_{qT}^2}\left(c^{\dag}c\right)^2\sigma_z.
\end{equation}
 \section{Derivation of the second-order correlation for the controlled system }
 In the limit of a weak driving field, the state $|\psi\rangle$ can be derived using perturbation theory. In order to show explicitly the second-order correlation,
up to second order in $|\varepsilon|$, we set $|\psi\rangle=C_{00}|00\rangle+C_{01}|01\rangle+C_{02}|02\rangle+C_{10}|10\rangle+C_{11}|11\rangle+C_{12}|12\rangle+C_{20}|20\rangle+C_{21}|21\rangle+C_{22}|22\rangle$.
The steady solution for the coefficients $C_{{n_a}{n_c}}$ can be given by the Schr\"{o}dinger equation for $|\psi\rangle$, when the limit of neglecting pure dephasing is neglected,
\begin{equation}
i\frac{{\rm d}|\psi\rangle}{{\rm d}t}=H_{\rm tot}|\psi\rangle \label{shordinger},
\end{equation}
assuming ${\rm d}\left|\psi(t)\right>/{\rm d}t=0$ , and
\begin{equation}
0=\Delta_s C_{10}-\varepsilon C_{11}+\frac{i}{2}\left(\sqrt{\kappa\gamma}-\sqrt{\kappa\gamma_f}\right)C_{01},
\end{equation}
\begin{equation}
0=\Delta_s C_{10}-\varepsilon C_{11}+\frac{i}{2}\left(\sqrt{\kappa\gamma}-\sqrt{\kappa\gamma_f}\right)C_{01},
\end{equation}
\begin{equation}
\begin{aligned}
0=&\left(\Delta_s+\Delta\right)C_{11}-\varepsilon C_{10}-\sqrt{2}\varepsilon C_{12}\\
&+\frac{i}{\sqrt{2}}\left(\sqrt{\kappa\gamma}-\sqrt{\kappa\gamma_f}\right)C_{02}-\frac{i}{\sqrt{2}}(\sqrt{\kappa\gamma}-\sqrt{\kappa\gamma_f})C_{20},
\end{aligned}
\end{equation}
\begin{equation}
\begin{aligned}
0=&\left(\Delta_s+2\Delta-2\chi\right) C_{12}-\sqrt{2}\varepsilon C_{11}\\
&-i\left(\sqrt{\kappa\gamma}-\sqrt{\kappa\gamma_f}\right)C_{21},
\end{aligned}
\end{equation}
\begin{equation}
\begin{aligned}
0=&-\varepsilon C_{00}-\sqrt{2}\varepsilon C_{02}-\frac{i}{2}\left(\sqrt{\kappa\gamma}-\sqrt{\kappa\gamma_f}\right)C_{10}+\Delta C_{01},
\end{aligned}
\end{equation}
\begin{equation}
\begin{aligned}
0=&\left(2\Delta-2\chi\right)C_{02}-\sqrt2\varepsilon C_{01}-\frac{i}{\sqrt{2}}(\sqrt{\kappa\gamma}-\sqrt{\kappa\gamma_f})C_{11},
\end{aligned}
\end{equation}
\begin{equation}
\begin{aligned}
0=&\left(2\Delta_s+\Delta\right)C_{21}-\varepsilon C_{20}-\sqrt2\varepsilon C_{22}\\
&+i\left(\sqrt{\kappa\gamma}-\sqrt{\kappa\gamma_f}\right)C_{12},
\end{aligned}
\end{equation}
\begin{equation}
\begin{aligned}
0=&2\Delta_s C_{20}-\varepsilon C_{21}+\frac{i}{\sqrt2}\left(\sqrt{\kappa\gamma}-\sqrt{\kappa\gamma_f}\right)C_{11},
\end{aligned}
\end{equation}
\begin{equation}
\begin{aligned}
0=&\left(2\Delta_s+2\Delta-2\chi\right)C_{22}-\sqrt2\varepsilon C_{21}.
\end{aligned}
\end{equation}
Due to the weak limit of the driving field, we can assume $C_{00}\to1$. And the equations are now closed (i.e., nine equations for nine parameters). Thus, it is possible to obtain the analytical solution of the system. However, the solution is cumbersome, but we can neglect the higher order terms in $|\varepsilon|$, obtaining
\begin{equation}\label{analytical g2a}
\begin{aligned}
P_1=&|C_{10}|^2+|C_{11}|^2+|C_{12}|^2\approx&\frac{|\left(\sqrt{\kappa\gamma}-\sqrt{\kappa\gamma_f}\right)\varepsilon|^2}{|\left(\Delta-\chi\right)\Delta_s|^2}\\
\end{aligned}
\end{equation}
\begin{equation}\label{analytical g2b}
\begin{aligned}
P_2=&|C_{20}|^2+|C_{21}|^2+|C_{22}|^2\\
\approx&\frac{|\left(\sqrt{\kappa\gamma}-\sqrt{\kappa\gamma_f}\right)\varepsilon|^4|\Delta_s-2\chi+\Delta|^2}{2|\left(\Delta-\chi\right)\left(\Delta-2\chi\right)\left(\Delta_s+\Delta-\chi\right)\Delta_s^2|^2}
\end{aligned}
\end{equation}
We substitute Eqs.~(\ref{analytical g2a})(\ref{analytical g2b}) into Eq.~(\ref{g2p}), and then we can derive the analytical solution of the $g^{(2)}(0)$ in the form of Eq.(\ref{analytical}).


\begin{thebibliography}{99}

\bibitem{Astrom} K. J. {\AA}str\"{o}m and R.M. Murray, {\it Feedback Systems: An Introduction for Scientists and Engineers} (Princeton University Press, Princeton, NJ, 2008).

\bibitem{Wiseman} H. M. Wiseman and G. J. Milburn, {\it Quantum Measurement and Control} (Cambridge University Press, Cambridge, England, 2009).
\bibitem{Dong} D.~Y. Dong and I.~R. Petersen, IET Control Theory and Applications {\bf 4}, 2651 (2010).
\bibitem{Altafini}C. Altafini and F. Ticozzi, IEEE Trans. Automat. Contr.
{\bf 57}, 1898 (2012).
\bibitem{Kerckhoff0} J. Kerckhoff, R. W. Andrews, H. S. Ku, W. F. Kindel, K. Cicak, R. W. Simmonds, and K. W. Lehnert, Phys. Rev. X {\bf 3}, 021013 (2013).

\bibitem{Belavkin} V. P. Belavkin, J. Multivariate Anal. {\bf 42}, 171 (1992);
 Commun. Math. Phys. {\bf 146}, 611 (1992);
 Theor. Probab. Appl. {\bf 38}, 573 (1993).



\bibitem{Wiseman2} H. M. Wiseman and G. J. Milburn, Phys. Rev. A {\bf 49}, 1350 (1994).

\bibitem{JGough1} J.~E. Gough and S. Wildfeuer, Phys. Rev. A {\bf 80}, 042107 (2009).
\bibitem{Iida0} S. Iida, M. Yukawa, H. Yonezawa, N. Yamamoto, and A. Furusawa,
IEEE Trans. Automat. Contr. {\bf 57}, 2045 (2012).

\bibitem{LKThomsen} L.~K. Thomsen, S. Mancini, and H.~M. Wiseman, Phys. Rev. A {\bf 65}, 061801 (2002).

\bibitem{JWang} J. Wang and H.~M. Wiseman, Phys. Rev. A {\bf 64}, 063810 (2001).
 \bibitem{Stockton} J.~K. Stockton, R. van Handel, and H. Mabuchi, Phys. Rev. A {\bf 70}, 022106
(2004).
\bibitem{Handel} R. van Handel, J.~K. Stockton, and H. Mabuchi, IEEE Trans.
Automat. Contr. {\bf 50}, 768 (2005).
\bibitem{Ibarcq} P. Campagne-Ibarcq, E. Flurin, N. Roch, D. Darson, P. Morfin, M. Mirrahimi, M. H. Devoret, F. Mallet, and B. Huard,
Phys. Rev. X {\bf 3}, 021008 (2013).

\bibitem{DVitali} D. Vitali, P. Tombesi, and G. J. Milburn, Phys. Rev. A {\bf 57}, 4930 (1998).
\bibitem{Ganesan} N. Ganesan and T.-J. Tarn, Phys. Rev. A {\bf 75}, 032323
(2007).
\bibitem{Jzhang} J. Zhang, R.-B. Wu, C.-W. Li, and T.-J. Tarn, IEEE Trans.
Automat. Contr. {\bf 55}, 619 (2010).
\bibitem{Fzhang} G.~F. Zhang and M.~R. James, IEEE Trans.
Automat. Contr. {\bf 56}, 1535 (2011).
\bibitem{Qi} B. Qi and L. Guo, Sys.
Contr. Lett. {\bf 59}, 333 (2010).

\bibitem{Vijay} R. Vijay, C. Macklin, D. H. Slichter, S. J. Weber, K. W. Murch, R. Naik, A. N. Korotkov and I. Siddiqi, Nature {\bf 490}, 77 (2012).

\bibitem{CAhn_and_ACDoherty} C. Ahn, A.~C. Doherty, and A.~J. Landahl, Phys. Rev. A {\bf 65}, 042301 (2002); J. Kerckhoff, H.~I. Nurdin, D.~S. Pavlichin, and H.
Mabuchi, Phys. Rev. Lett. {\bf 105}, 040502 (2010).
\bibitem{Xue} S.-B. Xue,
R.-B. Wu, W.-M. Zhang, J. Zhang, C.-W. Li, and T.-J. Tarn, Phys.
Rev. A {\bf 86}, 052304 (2012).

\bibitem{JWang_and_Wiseman} J. Wang, H.~M. Wiseman, and G.~J. Milburn, Phys. Rev. A {\bf 71}, 042309
(2005).
\bibitem{Yamamoto} N. Yamamoto, K. Tsumura and S. Hara, Automatica, {\bf
43}, 981 (2007).
\bibitem{zLiu} Z. Liu, L.~L. Kuang, K. Hu, L.~T. Xu, S.~H.
Wei, L.~Z. Guo, and X.~Q. Li, Phys. Rev. A {\bf 82}, 032335
(2010).

\bibitem{AHopkins_and_KJacobs} A. Hopkins, K. Jacobs, S. Habib, and K.~C. Schwab, Phys. Rev. B {\bf 68}, 235328
(2003).
\bibitem{jzhangl} J. Zhang, Y.-X. Liu, and F. Nori, Phys. Rev. A {\bf 79},
052102 (2009).
\bibitem{Greenberg} Ya. S. Greenberg, E. Il¡¯ichev, and F. Nori, Phys.
Rev. B {\bf 80}, 214423 (2009); K. Xia and J. Evers, Phys. Rev. B
{\bf 82}, 184532 (2010).
\bibitem{Woolley} M.~J. Woolley, A.~C. Doherty, and G. J.
Milburn, Phys. Rev. B {\bf 82}, 094511 (2010).

\bibitem{JohnComes} J. Combes and K. Jacobs, Phys. Rev. Lett. {\bf 96}, 010504 (2006).
J. Combes, H. M. Wiseman, and K. Jacobs, Phys. Rev. Lett. {\bf
100}, 160503 (2008).
\bibitem{jzhangl2} J. Zhang, Y.-X. Liu, R.-B. Wu, C.-W. Li, and
T.-J. Tarn, Phys. Rev. A {\bf 82}, 022101 (2010).

\bibitem{Wisman1} H. M. Wiseman and G. J. Milburn, Phys. Rev. Lett. {\bf 70}, 548 (1993); H. M. Wiseman, Phys. Rev. A {\bf 49}, 2133 (1994).

\bibitem{Mabuchi} H. Mabuchi and A. C. Doherty, Science {\bf 298}, 1372 (2002).

\bibitem{Doherty} A. C. Doherty, S. Habib, K. Jacobs, H. Mabuchi, and S. M. Tan, Phys. Rev. A {\bf 62}, 012105 (2000).

\bibitem{Mancini} S. Mancini, Phys. Rev. A {\bf 73}, 010304(R) (2006).

\bibitem{Wisman4} H. M. Wiseman and G. J. Milburn, Phys. Rev. A {\bf 49}, 4110 (1994).

\bibitem{Lloyd} S. Lloyd, Phys. Rev. A {\bf 62}, 022108 (2000).

\bibitem{James} M. R. James, H. I. Nurdin, and I. R. Petersen, IEEE Trans. Automat. Contr. {\bf 53}, 1787 (2008).

\bibitem{Yanagisawa}M. Yanagisawa, H. Kimura, IEEE Trans. Automat. Contr. {\bf 48}, 2107 (2003).

\bibitem{Gough} J. Gough and M.~R. James, IEEE Trans. Automat.Contr. {\bf 54}, 2530 (2009).

\bibitem{Riste} D. Riste, C. C. Bultink, K. W. Lehnert, and L. DiCarlo, Phys. Rev. Lett. {\bf 109}, 240502 (2012).

\bibitem{Sciarrino} F. Sciarrino, M. Ricci, F. De Martini, R. Filip, and L. Mi\v{s}ta, Jr., Phys. Rev. Lett. {\bf 96}, 020408 (2006).

\bibitem{Brakhane} S. Brakhane, W. Alt, T. Kampschulte, M. M. Dorantes, R. Reimann, S. Yoon, A. Widera, and D. Meschede, Phys. Rev. Lett. {\bf 109}, 173601 (2012).

\bibitem{Yonezawa} H. Yonezawa, D. Nakane, T. A. Wheatley, K. Iwasawa, S. Takeda, H. Arao, K. Ohki, K. Tsumura, D. W. Berry, T. C. Ralph, H. M. Wiseman, E. H. Huntington, and A. Furusawa, Science {\bf 337}, 1514 (2012).

\bibitem{Gieseler} J. Gieseler, B. Deutsch, R. Quidant, and L. Novotny, Phys. Rev. Lett. {\bf 109}, 103603 (2012).

\bibitem{Berry0} D. W. Berry and H. M. Wiseman, Phys. Rev. Lett. {\bf 85}, 5098 (2000).

\bibitem{Kubanek} A. Kubanek, M. Koch, C. Sames, A. Ourjoumtsev, P. W. H. Pinkse, K. Murr and G. Rempe, Nature {\bf 462}, 898 (2009).

\bibitem{Sayrin} C. Sayrin,  I. Dotsenko, X. Zhou, B. Peaudecerf,    T. Rybarczyk,   S. Gleyzes, P. Rouchon, M. Mirrahimi, H. Amini, M. Brune, J. M. Raimond, and S. Haroche, Nature {\bf 477}, 73 (2011).

\bibitem{Zhou} X. Zhou, I. Dotsenko, B. Peaudecerf, T. Rybarczyk, C. Sayrin, S. Gleyzes, J. M. Raimond, M. Brune, and S. Haroche, Phys. Rev. Lett. {\bf 108}, 243602 (2012).

\bibitem{Nelson} R. J. Nelson, Y. Weinstein, D. Cory, and S. Lloyd, Phys. Rev. Lett. {\bf 85}, 3045 (2000).

\bibitem{Mabuchi1} H. Mabuchi, Phys. Rev. A {\bf 78}, 032323 (2008).

\bibitem{Zhou0} Z .F. Zhou, C. J. Liu, Y. M. Fang, J.Zhou, R. T. Glasser, L. Q. Chen, J. T. Jing, and W. P. Zhang, Appl. Phys. Lett. {\bf 101}, 191113 (2012).

\bibitem{Kerckhoff} J. Kerckhoff, and K. W. Lehnert, Phys. Rev. Lett. {\bf 109}, 153602 (2012).

\bibitem{Iida}S. Iida, M. Yukawa, H. Yonezawa, N. Yamamoto, and A. Furusawa, IEEE Trans. Automat. Contr. {\bf 57}, 2045 (2012).

\bibitem{Jaksch}D. Jaksch, C. Bruder, J. I. Cirac, C. W. Gardiner, and P. Zoller, Phys. Rev. Lett. {\bf 81}, 3108 (1998).

\bibitem{Birnbaum} K. M. Birnbaum, A. Boca, R. Miller, A. D. Boozer, T. E. Northup, and H. J. Kimble, Nature {\bf 436}, 87 (2005).

\bibitem{Fink} J. M. Fink, M. Goppl, M. Baur, R. Bianchetti, A. Leek,P. J. Blais, and A. Wallraff, Nature {\bf 454}, 315 (2008).

\bibitem{Bishop} L. S. Bishop, J. M. Chow, J. Koch, A. A. Houck, M. H. Devoret, E. Thuneberg, S. M. Girvin, and R. J. Schoelkopf, Nat. Phys. {\bf 5}, 105 (2009).

\bibitem{Politi} A. Politi, J.C.F. Matthews, and J. L. O'Brien, Science {\bf 325}, 1221 (2009).

\bibitem{Matthews} J.C.F. Matthews, A. Politi, A. Stefanov, and J. L. O'Brien, Nat. Photonics {\bf 3}, 346 (2009).

\bibitem{Berry} D. W. Berry and H. M. Wiseman, Nat. Photonics {\bf 3}, 317 (2009).

\bibitem{J.Zhang} J. Zhang, R.-B. Wu , Y.-X. Liu, C.-W Li, and T.-J. Tarn, IEEE Trans. Automat. Contr. {\bf 57}, 1997 (2012).


\bibitem{You} J. Q. You, Y.-X. Liu, C. P. Sun, and F. Nori, Phys. Rev. B {\bf 75}, 104515 (2007).

\bibitem{Hoffman} A. J. Hoffman, S.J. Srinivasan, S. Schmidt, L. Spietz, J. Aumentado, H. E. T\"{u}reci, and A. A. Houck, Phys. Rev. Lett. {\bf 107}, 053602 (2010).

\bibitem{Nation} P. D. Nation, and M. P. Blencowe, E. Buks, Phys. Rev. B {\bf 78}, 104516 (2008).

\bibitem{Tornes} I. Tornes, and D. Stroud, Phys. Rev. B {\bf 77}, 224513 (2008).

\bibitem{Buluta0} I. Buluta, S. Ashhab, and F. Nori, Rep. Prog. Phys. {\bf 74}, 104401 (2011).
\bibitem{Georgescu} I. Georgescu, and F. Nori, Phys. World {\bf 25}, 16-17 (2012).
\bibitem{Buluta1} I. Buluta, and F. Nori, Science {\bf 326}, 108-111 (2009).
\bibitem{Nation1} P. D. Nation, J. R. Johansson, and M. P. Blencowe, and F. Nori, Rev. Mod. Phys. {\bf 84}, 1-24 (2012).
\bibitem{Youf0} J. Q. You, and F. Nori, Phys. Today {\bf 58}, 42-47 (2005).
\bibitem{Youf1} J. Q. You, and F. Nori, Nature {\bf 474}, 589 (2011).
\bibitem{Xiang0} Z.-L. Xiang, S. Ashhab, J. Q. You, and F. Nori, Rev. Mod. Phys. {\bf 85}, 623 (2013).

\bibitem{Gardiner} C.~W. Gardiner and M.~J. Collett, Phys. Rev. A {\bf 31}, 3761 (1985); C.~W. Gardiner and P. Zoller,
{\it Quantum Noise} (Springer-Verlag, Berlin, 2004) (3rd edition).

\bibitem{Jacobs} A detailed and pedagogical derivation of the input-output formalism can be found in: K. Jacobs, PhD dissertation, Imperial, London, Eprint:arXiv:quant-ph/9810015.

\bibitem{G. Wendin} G. Wendin and V.~S. Shumeiko, {\it Handbook of Theoretical and Computational Nanotechnology} edited by M. Rieth and W. Schommers (ASP, Los Angeles, 2006), Vol. {\bf 3}, pp. 223¨C309.

\bibitem{Beltran} M. A. Castellanos-Beltran, K. D. Irwin, G. C. Hilton, L. R. Vale, and K. W. Lehnert, Nature phys. {\bf 4}, 929 (2008).

\bibitem{Liu1} Y.-X. Liu, A. Miranowicz, Y. B. Gao, J. Bajer, C. P. Sun, and F. Nori, Phys. Rev. A {\bf 82}, 032101 (2010).

\bibitem{Gibbs} H. M. Gibbs, S. L. McCall, and T. N. C. Venkatesan, Phys. Rev. Lett. {\bf 36}, 1135 (1976).

\bibitem{Rempe} G. Rempe, R. J. Thompson, R. J. Brecha, W. D. Lee, and H. J. Kimble, Phys. Rev. Lett. {\bf 67}, 1727 (1991).

\bibitem{Sauer} A. Sauer, K. M. Fortier, M. S. Chanf, C. D. Hamley, and M. S. Chapman, Phys. Rev. A {\bf 69}, 051804 (2004).

\bibitem{Gupta} S. Gupta, K. L. Moore, K. W. Murch, and D. M. Stamper-Kurn, Phys. Rev. Lett. {\bf 99}, 213601 (2007).

\bibitem{Brennecke} F. Brennecke, S. Ritter, T. Donner, and T. Esslinger, Science {\bf 322}, 235 (2008).

\bibitem{Xiao} Y. F Xiao, \c{S}. K. \"{O}zdemir, V. Gaddam, C. H. Dong, N. Imoto, and L. Yang, Opt. Express {\bf 16}, 21462
(2008).

\bibitem{Hui Wang} H. Wang, H. C. Sun, J. Zhang, and Y.-X. Liu, Science China {\bf 55}, 2264 (2012).

\bibitem{Puri}R. R. Puri, {\it Mathematical Methods of Quantum Optics} (Springer-Verlag, Berlin, 2001).

\bibitem{Adam} A. Miranowicz, M. Paprzycka, Y.-X. Liu, J. Bajer, and F. Nori, Phys. Rev. A {\bf 87}, 023809 (2013).

\bibitem{Xu} X. W. Xu, and Y. J. Li, J. Phys. B: At. Mol. Opt. Phys. {\bf 46}, 035502 (2013).
\end{thebibliography}
\end{document}